\documentclass[11pt,english,superscriptaddress,showpacs,aps,preprint]{revtex4}

\usepackage[latin9]{inputenc}

\usepackage{amsmath}
\usepackage{amssymb}
\usepackage{graphicx}
\usepackage{color}

\makeatletter

\usepackage{slashed}

\def\as{a\!\!\!/}
\def\ks{k\!\!\!/}

\def\bs{b\!\!\!/}

\def\ds{\partial\!\!\!/}
\newcommand{\be}{\begin{equation}}
\newcommand{\ee}{\end{equation}}
\newcommand{\en}{\end{equation}}
\newcommand{\ba}{\begin{eqnarray}}
\newcommand{\ea}{\end{eqnarray}}
\newcommand{\bea}{\begin{eqnarray}}
\newcommand{\eea}{\end{eqnarray}}
\newcommand{\pa}{\partial}
\newcommand{\bq}{\begin{eqnarray}}
\newcommand{\eq}{\end{eqnarray}}

\makeatother

\usepackage{babel}

\begin{document}

\title {Effective potential in Lorentz-breaking field theory models}

\author{A. P. Baeta Scarpelli}
\email{scarpelli.abps@dpf.gov.br}
\affiliation{Centro Federal de Educa\c{c}\~ao Tecnol\'ogica - MG \\
Avenida Amazonas, 7675 - 30510-000 - Nova Gameleira - Belo Horizonte
-MG - Brazil}
\affiliation{Setor T\'{e}cnico-Cient\'{\i}fico - Departamento de Pol\'\i cia Federal \\
Rua Nascimento Gurgel, 30 - Gutierrez - Belo Horizonte - MG - Brazil}

\author{L. C. T. Brito}
\email{lcbrito@dfi.ufla.br}
\affiliation{Departamento de Física, Universidade Federal de Lavras, Caixa Postal 3037,
37200-000, Lavras, MG, Brasil}

\author{J. C. C. Felipe}
\email{jean.cfelipe@ufvjm.edu.br}
\affiliation{Departamento de Física, Universidade Federal de Lavras, Caixa Postal 3037,
37200-000, Lavras, MG, Brasil}
\affiliation{Instituto de Engenharia, Ciência e Tecnologia, Universidade Federal dos Vales do Jequitinhonha e Mucuri, Avenida Manoel Bandeira, 460 - 39440-000 - Veredas - Janaúba - MG- Brazil}

\author{J. R. Nascimento}
\email{jroberto@fisica.ufpb.br}
\affiliation{Departamento de Física, Universidade Federal da Paraíba, Caixa Postal 5008,
58051-970 João Pessoa, Paraíba, Brazil}

\author{A. Yu. Petrov}
\email{petrov@fisica.ufpb.br}
\affiliation{Departamento de Física, Universidade Federal da Paraíba, Caixa Postal 5008,
58051-970 João Pessoa, Paraíba, Brazil}


\begin{abstract}
We calculate explicitly the one-loop effective potential in different Lorentz-breaking field theory models. First, we consider a Yukawa-like theory and, then, some examples of Lorentz-violating extensions of scalar QED. We observed, for the extended QED models, that the resulting effective potential converges to the known result in the limit in which Lorentz-symmetry is restored. Besides, the one-loop corrections to the effective potential in all the cases we studied depend on the background tensors responsible for the Lorentz symmetry violation. This have consequences in physical quantities like, for example, in the induced mass due to Coleman-Weinberg mechanism.
\end{abstract}

\maketitle

\section{Introduction}
\label{I}

The hypothesis of Lorentz symmetry breaking and its possible impacts within different field theory models are now intensively discussed. Many examples of Lorentz-breaking extensions of well-known models are presented in \cite{Kostel}. These theories are known to display nontrivial features both at classical \cite{class,CMR} (birefringence of electromagnetic waves in the vacuum, superluminal modes of propagation, non-zero magnetic moment for neutral elementary particles, etc) and quantum levels. Concerning quantum effects, one of the most important directions of study is the investigation whether new Lorentz-breaking terms are induced starting from some underlying coupling which breaks this symmetry. The new generated extensions are, thus, treated as emergent phenomena, like in the seminal paper \cite{JK} in which it was shown that the famous Carroll-Field-Jackiw term \cite{CFJ} arises as a quantum correction. Following this procedure, many new examples of additive terms were obtained, such as the four-dimensional gravitational Chern-Simons, the aether-like and higher-derivative Lorentz-breaking terms \cite{ourHD}.

In the study of quantum corrections, sometimes it is necessary to understand the low-energy dynamics. In this context, the key tool for this investigation is the effective potential \cite{Col-Wei}. In this interesting approach, also know as the Coleman-Weinberg mechanism, the Higgs potential is induced by radiative corrections, in the place of being inserted by hand. Higher-loop graphs are considered in order to generate an effective potential, which may produce spontaneous symmetry breaking. Certainly, it would be interesting to find the possible Lorentz-breaking modifications of the effective potential in different Lorentz-breaking extensions of known field theory models. Up to now, this study has been performed only for a purely scalar theory \cite{aether}. Therefore, the natural continuation would consist in studying the effective potential in other Lorentz-breaking field theories. In this paper, we carry out a study of effective potentials in Lorentz-breaking extensions of QED and of a Yukawa-like theory.

The paper is organized as follows: in section II, we set a Yukawa-like model which breaks Lorentz symmetry and, then, proceed to the calculation of the effective potential, obtaining the corrections depending on the Lorentz-breaking parameters; in section III, we perform the same study in the extended scalar QED for different kinds of Lorentz violation; we present our conclusions in section IV.

\section{Yukawa-like Lorentz-breaking theory}
\label{II}

We start our study with a Lorentz-breaking generalization of the Yukawa theory. The first attempt to study such  theory at the quantum level has been performed in \cite{aether}, where the following action in a $d$-dimensional space-time has been considered:
\bea
S_{1,Yu}=\int d^dx\bar{\psi}(i\ds-m+\as\phi)\psi.
\eea
However, while the derivative-dependent corrections in this theory (in particular, the aether ones) are nontrivial \cite{aether}, it is clear that the one-loop effective potential in this theory vanishes. Indeed, the effective potential, by definition, is evaluated at constant values of the background scalar field, that is, $\phi=\Phi$, with $\Phi$ a constant. When the scalar field is purely external, we have
\bea
\Gamma^{(1)}_1=-i{\rm Tr}\int\frac{d^dk}{(2\pi)^d}\ln(\ks+\as\Phi-m),
\eea
which, after a trivial change of variable, $k_m\to k_m+a_m\Phi$, becomes field-independent and, hence, vanishes. Therefore, we introduce another coupling, so that
\bea
S_{2,Yu}=\int d^dx\bar{\psi}(i\ds-\bs\gamma_5\phi)\psi,
\eea
in which $b_m$ is a pseudovector and the mass is taken to be zero. The corresponding one-loop effective potential is given by
\bea
\Gamma^{(1)}_2=-i{\rm Tr}\int\frac{d^dk}{(2\pi)^d}\ln(\ks-\bs\gamma_5\Phi).
\eea
So, we should calculate the trace. It is well known that to do it, we should rewrite this expression in terms of some quadratic operator. Let us restrict ourselves to the usual case $d=4$.

First of all, by dimensional reasons and due to the properties of the trace, the one-loop effective potential is even in $\Phi$ and $\bs$. So, we can use the symmetrized form of the integral,
\bea
\Gamma^{(1)}_2=-\frac{i}{2}\left\{{\rm Tr}\int\frac{d^dk}{(2\pi)^d}\left[\ln(\ks-\bs\gamma_5\Phi)+\ln(\ks+\bs\gamma_5\Phi)\right]
\right\},
\eea
which yields
\bea
\Gamma^{(1)}_2=-\frac{i}{2}\left\{{\rm Tr}\int\frac{d^dk}{(2\pi)^d}\ln\left(k^2-b^2\Phi^2+2(b\cdot k)\Phi \gamma_5\right)
\right\}.
\eea
Now, we face the problem of calculating the matrix trace ${\rm Tr}\ln\left[C+(A+B\bs)\gamma_5\right]$, where $C$, $A$ and $B$ are some $c$-numbers. One can easily show that
\bea
{\rm Tr}\ln\left[C+(A+B\bs)\gamma_5\right]=D\left\{\ln C+\frac{1}{2}\ln\left(1-\frac{A^2-B^2b^2}{C^2}\right)\right\},
\eea
where $D$ is the dimension of the Dirac matrices in the corresponding representation. In our case, with $D=4$, it is obtained in the form:
\bea
\Gamma^{(1)}_2=-2i{\rm Tr}\int\frac{d^dk}{(2\pi)^d}\left\{\ln(k^2-b^2\Phi^2)+\frac{1}{2}\ln\left[1-\frac{4(b\cdot k)^2\Phi^2}{(k^2-b^2\Phi^2)^2}\right]
\right\}.
\eea
For calculating this integral, we first perform the replacement, $k^ak^b\to\frac{1}{4}k^2\eta^{ab}$, and, then, we carry out the Wick rotation $k_0\to ik_{0E}$, which, for our signature $(+---)$, yields $k^2\to -k^2_E$. As the result, we get
\bea
\Gamma^{(1)}_2=2{\rm Tr}\int\frac{d^dk_E}{(2\pi)^d}\left\{\ln(k^2_E+b^2\Phi^2)+\frac{1}{2}\ln\left[1+
\frac{b^2k^2_E\Phi^2}{(k^2_E+b^2\Phi^2)^2}\right]
\right\},
\eea
which yields, after some simplifications,
\bea
\Gamma^{(1)}_2=\int\frac{d^dk_E}{(2\pi)^d}\ln\Big[(k^2_E+b^2\Phi^2)^2+ k^2_E b^2\Phi^2
\Big].
\eea
If we use
\bea
(k^2_E+b^2\Phi^2)^2+k^2_E b^2\Phi^2=(k^2_E+r_1 \Phi^2)(k^2_E+r_2 \Phi^2),
\eea
with
\bea
r_{1,2}=- \frac {b^2}{2}\left(3 \pm \sqrt{5}\right),
\eea
and integration with use of Dimensional Reduction, we obtain
\bea
\label{epsc}
\Gamma^{(1)}_2&=&\mu^{-\epsilon}\int\frac{d^dk_E}{(2\pi)^d}\ln\left[(k^2_E+r_1\Phi^2)(k^2_E+r_2\Phi^2)\right] \nonumber \\
&=&\mu^{-\epsilon}\frac{\Gamma(\epsilon/2)}{16\pi^2(-1+\epsilon/2)(2+\epsilon/2)}
\left[\left(\frac{r_1\Phi^2}{4\pi}\right)^{2+\epsilon/2}+\left(\frac{r_2\Phi^2}{4\pi}\right)^{2+\epsilon/2}\right]\nonumber\\
&=&-\frac{\Phi^4}{16\pi^2}\left\{7 b^4\left[\frac{1}{\epsilon}
+ \frac{1}{2} \ln\left(\frac{\Phi^2}{\mu^2}\right)\right]
+\frac 12\left(r^2_1\ln r_1+r^2_2\ln r_2\right)\right\},
\eea
where $\epsilon=d-4$. We see that this result, first, is quartic in the Lorentz-breaking vector (and hence very small) and, second, involves a logarithmic dependence on the Lorentz-breaking parameter.
It is necessary to comment here the on fine-tuning problem \cite{Collins}, that is, on the possibility of large Lorentz-breaking quantum corrections. It was shown in \cite{Collins} that in certain cases the loop corrections are not suppressed if the Lorentz-breaking parameters are small, and hence they should essentially affect the effective dynamics. While in certain cases the large corrections are really observed, see f.e. \cite{Reyes}, it is not so in our case. Indeed, in (\ref{epsc}), the effective potential depends on the Lorentz-breaking vector $b_a$ as $b^4\ln b^2$ which goes to zero at $b_a\to 0$. This is reasonable since our Lorentz-breaking term does not affect the propagators but only couplings, and, moreover, our vertex vanishes at $b_a\to 0$.

\section{Lorentz-violating electrodynamics}\label{s1}

Let us now study the effective potential for a more elaborated model. We consider the following Lorentz-breaking extension of scalar QED,
\be\label{1}
{\cal L}= -\frac{1}{4}F_{\mu \nu}F^{\mu \nu}+\frac{\xi}{2}\epsilon^{\mu \nu \alpha \beta}c_\mu A_\nu \pa_\alpha A_\beta -\frac{\alpha}{2}b^\mu F_{\mu \nu}b_\alpha F^{\alpha \nu} +\kappa_{\mu \nu}(D^\mu \phi)^*D^\nu \phi -\frac{\lambda}{4!}(\phi^* \phi)^2,
\ee
in which, as usual, $D_\mu\phi=(\pa_\mu-ieA_\mu)\phi$ and $F_{\mu \nu}=\pa_\mu A_\nu - \pa_\nu A_\mu$. The constant tensors $c_\mu$, $b_\mu$ and $\kappa_{\mu \nu}$ are responsible for the violation of Lorentz symmetry. The breaking terms are controlled by the dimensionless parameters $\xi$ and $\alpha$, which from now on will be used to turn off some terms and simplify the analysis.

\subsection{QED with the Carroll-Field-Jackiw term}
\subsubsection{Classical action}

If we set $\xi=1$, $\alpha=0$ and $\kappa^\mu_\nu=\delta^\mu_\nu$, the unique Lorentz-breaking term which remains is the CPT-odd one, known as the Carroll-Field-Jackiw term. The Lagrangian density is then given by
\begin{equation}\label{2}
{\cal L}_1=-\frac{1}{4}F_{\mu \nu}F^{\mu \nu}+\frac{1}{2}\epsilon^{\mu \nu \alpha \beta}c_\mu A_\nu \pa_\alpha A_\beta+(D_\mu \phi)^*D^\mu \phi
-\frac{\lambda}{4!}(\phi^* \phi)^2.
\end{equation}

 In order to perform the calculation of the effective potential, we make use of the background field method. For this, we write the scalar field as $\phi \rightarrow \phi + \Phi$, in which $\Phi$ is a constant background scalar field. In addition, we decompose the complex scalar field in terms of real scalar fields as $\phi=\frac{1}{\sqrt{2}}(\phi_{1}+i\phi_{2})$. The quadratic part of the action is then rewritten as
\begin{eqnarray}\label{5}
\int d^4x{\cal L}(\phi_a,\Phi_a,A^{\mu})&=&\int d^4xd^4y\Big\{\frac{1}{2}A^{\mu}(x)(i\Delta^{-1})_{\mu\nu}A^{\nu}(y) +\nonumber \\&&
\frac{1}{2}\phi^{a}(x)(i{\cal D}^{-1})_{ab}\phi^{b}(y)+A^{\mu}(x){\cal M}_{\mu a}\phi^{a}(y)\Big\},
\end{eqnarray}
with
\begin{eqnarray}\label{6}
(\Delta^{-1})^{\mu\beta}&=&(\Box +e^{2}\Phi^{2})\eta^{\mu\beta}-\pa^{\mu}\pa^{\beta}(1-\chi)+\epsilon^{\nu\mu\alpha\beta}c_{\nu}\pa_{\alpha}, \\
({\cal D}^{-1})^{ab}&=& \left(-\Box -\frac{\lambda}{6}\Phi^{2} \right)\delta^{ab}-\frac{\lambda}{3}\Phi^{a}\Phi^{b}\\
\mbox{and} \,\,\,\,\,\,\,\,
{\cal M}_{\mu a}&=& -e\epsilon_{ab}\Phi^{b}\partial_\mu.
\end{eqnarray}
In the equations above, the Latin indices refer to the real components of the scalar field and $\chi$ is a gauge parameter. We also note that the inverse of the propagators are written in function of the background field $\Phi$.

\subsubsection{The Effective Potential}
We proceed to the calculation of the effective potential following the lines of the paper from Jackiw \cite{Jackiw}. The effective potential can be written as
\begin{eqnarray}\label{7}
\Gamma(\Phi)&=&-\frac{i}{2}\int \frac{d^{d}k}{(2\pi)^{d}}\ln \det(i{\cal D}^{-1})\nonumber \\&& - \frac{i}{2}\int \frac{d^{d}k}{(2\pi)^{d}}\ln \det\big[i\Delta^{-1}+iN\big],
\end{eqnarray}
in which, we have used the identity $\ln\det A= Tr\ln A$ and
\begin{equation}
N^{\mu\beta}={\cal M}^{\mu a}{\cal D}_{ab}{\cal M}^{\beta b}.
\end{equation}

Lets start with the determinant of the inverse propagator of the scalar field. In momentum space, the propagator is given by
\be\label{10}
{\cal D}_{ab}= \frac{\Phi_{a}\Phi_{b}}{\Phi^{2}}\frac{i}{k^{2}-\frac{\lambda}{2}\Phi^{2}}+ \Bigg(\delta_{ab}-\frac{\Phi_{a}\Phi_{b}}{\Phi^{2}}\Bigg)\frac{i}{k^{2}-\frac{\lambda}{6}\Phi^{2}},
\ee
and we have, for the determinant,
\begin{equation}\label{11}
\det ({\cal D}^{-1})=\Bigg({k^{2}-\frac{\lambda}{6}\Phi^{2}}\Bigg)\Bigg({k^{2}-\frac{\lambda}{2}\Phi^{2}}\Bigg).
\end{equation}

For the second term of eq. (\ref{7}), we first obtain
\begin{equation}
N_{\mu\beta}=\frac{ie^{2}{\Phi}^{2} k^{\mu}k^{\beta}}{k^{2}-\frac{\lambda {\Phi}^{2}}{6}}
\end{equation}
and, then, the sum of terms
\begin{eqnarray}\label{14}
i(\Delta^{-1})^{\mu\beta}+iN^{\mu\beta}&=&\Bigg(-k^{2}+e^{2}\Phi^{2}\Bigg)\eta^{\mu\beta}-
\frac{e^{2}\Phi^{2}k^{\mu}k^{\beta}}{\Big(k^{2}-\frac{\lambda\Phi^{2}}{6}\Big)}+(1-\chi)k^{\mu}k^{\beta}\nonumber \\
&-& i\epsilon^{\nu\mu\alpha\beta}c_{\nu}k_{\alpha}.
\end{eqnarray}
After a lengthy calculation, one obtains
\begin{eqnarray}\label{15}
\det\big[i\Delta^{-1}+iN\big]&=&-a \Big(a+g k^2\Big)\Big(a^2+c^2 k^2-(c\cdot k)^2\Big),
\end{eqnarray}
with
\be \label{16}
a= -k^{2}+e^{2}\Phi^{2}
\ee
and
\be \label{16a}
g = (1-\chi)-\frac{e^{2}\Phi^{2}}{k^{2}-\frac{\lambda\Phi^{2}}{6}}
\ee

We, thus, stay with
\bq \label{17}
\Gamma(\Phi)&=&-\frac{i}{2}\int \frac{d^{d}k}{(2\pi)^{d}}\ln\Big[\Big({k^{2}-\frac{\lambda}{6}\Phi^{2}}\Big)\Big({k^{2}-\frac{\lambda}{2}\Phi^{2}}\Big)\Big] \nonumber \\
&-&\frac{i}{2}\int \frac{d^{d}k}{(2\pi)^{d}}\ln\Big[-a \Big(a+g k^2\Big)\Big(a^2+c^2 k^2-(c\cdot k)^2\Big)\Big] \nonumber \\
&\equiv & \Gamma_1(\Phi)+\Gamma_2(\Phi) .
\eq
The first term is independent on the Lorentz-breaking vector and gives, with the help of the Dimensional Reduction procedure, the following result:
\be
\Gamma_1(\Phi)=-\frac{5}{288 \pi^2}\lambda^2 \Phi^4\left\{\frac{1}{\epsilon}+\frac 12 \left[\ln\left(\frac{\Phi^2}{\mu^2}\right)
+\ln\left(\frac{\lambda}{2}\right) -\frac{1}{10}\ln 3\right]\right\}
\ee
The second term, which involves the gauge and mixed propagators, will depend on the constant vector $c_\mu$. First, we can rewrite this term as
\begin{equation}
\Gamma_2(\Phi)=\frac{1}{2}\int \frac{d^{d}k_{E}}{(2\pi)^{d}}\ln\Bigg[\frac{(k^2_E+e^2\Phi^2)(k^{2}_{E}+c_1 \Phi^2)(k^{2}_{E}+c_{2}\Phi^2)
(k^{2}_{E}+a_{1})(k^{2}_{E}+a_{2})}{(k^{2}_{E}+\frac{\lambda \Phi^2}{6})}\Bigg],
\end{equation}
where
\be
c_{1,2}=-\frac{\lambda}{12}\left(1 \pm \sqrt{1 - \frac{24e^2}{\lambda}}\right)
\ee
and
\be
a_{1,2}=\frac{1}{8}\left\{-8 e^2 \Phi^2 + 3 c^2 \pm \sqrt{3 c^2\left[3 c^2 - 8 e^2 \Phi^2\right]}\right\},
\ee
in which we have used $k_\mu k_\nu \to \eta_{\mu \nu}k^2/4$. Again, we follow the procedures of dimensional reduction and obtain
\bq
&&\Gamma_2(\Phi)=-\frac{1}{32 \pi^2} \Phi^4 \left\{\left(3e^4 -\frac {\lambda}{3}e^2\right)\frac{1}{\epsilon} +\frac 12 \left(e^4 - \frac{\lambda}{3}e^2\right)
\ln\left(\frac{\Phi^2}{\mu^2}\right)+ \frac{c_1^2}{2} \ln (c_1) + \frac{c_2^2}{2} \ln (c_2)\right.\nonumber \\
&& \left. + \frac{e^4}{2} \ln(e^2) + \frac{\lambda^2}{72}
\ln\left(\frac{\lambda}{6}\right) \right\} \nonumber \\
&& -\frac{1}{32 \pi^2} \left\{ \left(-3c^2e^2 \Phi^2+\frac{9}{16}c^4\right) \frac{1}{\epsilon}+ \frac{a_1^2}{2} \ln \left(\frac{a_1}{\mu^2}\right)
+ \frac{a_2^2}{2} \ln \left(\frac{a_2}{\mu^2}\right)\right\}.
\eq
One should observe that we have chosen the gauge with $\chi=1$ (Feynman gauge), $c^2=c_\mu c^\mu$ and that the mass parameter $\mu^2$ appears as a feature of the dimensional reduction. It is also to be noted that only the last line is $c_\mu$-dependent and that, in the limit $c_\mu \to 0$, the result of Coleman and Weinberg \cite{Col-Wei} is recovered.

\subsection{CPT-even Lorentz-violating QED}\label{s2}
\subsubsection{Formulation of the model}

If we set $\xi=0$, $\alpha=1$ and $\kappa^\mu_\nu=\delta^\mu_\nu$, the unique Lorentz-breaking term which remains is the CPT-even one of the gauge sector, known as the aether term. The Lagrangian density is then given by
\begin{eqnarray}\label{19}
{\cal L}_2&=&-\frac{1}{4}F_{\mu \nu}F^{\mu \nu}-\frac{1}{2}b^\alpha F_{\alpha \mu}b_\beta F^{\beta \mu}+(D_\mu \phi)^*D^\mu \phi-\frac{\lambda}{4!}(\phi^* \phi)^2.
\end{eqnarray}

All the steps carried out in the first model, concerning the decomposition of the complex scalar field and the use of the background field method, are again applied, so that we get, for the quadratic part of the action,
\begin{eqnarray}\label{21}
\int d^4x{\cal L}(\phi_{a},\Phi_{a},A^{\mu})&=&\int d^4xd^4y\Big\{\frac{1}{2}A^{\mu}(x)(i\Delta^{-1})_{\mu\nu}A^{\nu}(y) +\nonumber \\&&
\frac{1}{2}\phi^{a}(x)(i{\cal D}^{-1})_{ab}\phi^{b}(y)+A^{\mu}(x){\cal M}_{\mu a}\phi^{a}(y)\Big\},
\end{eqnarray}
for which, in momentum-space, we have
\begin{eqnarray}\label{22}
(\Delta^{-1})_{\mu\beta}&=&\left[-k^{2}-(k \cdot b)^{2}+e^2 \Phi^2\right]\eta_{\mu\beta}+k_{\mu}k_{\beta}(1-\chi)+(k \cdot b)\Big(k_{\mu}b_{\beta}+k_{\beta}b_{\mu}\Big) -k^{2}b_{\beta}b_{\mu},\nonumber \\
({\cal D}^{-1})_{ab}&=&\left(k^2-\frac{\lambda}{6}\Phi^2 \right)\delta_{ab}-\frac{\lambda}{3}\Phi_{a}\Phi_{b}\nonumber \\
\mbox{and} \;\;\;\;   {\cal M}_{\mu a}&=& -ie\epsilon_{ab}\Phi^{b}k_\mu,
\end{eqnarray}
with $\chi$ again the gauge parameter.

\subsubsection{The effective potential}

We proceed, as before, to the calculation of the effective potential. The first step is the evaluation of the determinants. The scalar propagator, ${\cal D}^{ab}$, is the same as in the first model. So, we concentrate our efforts in the term which involves the gauge and the mixed sectors, for which we obtain
\be\label{25}
\det\big[i\Delta^{-1}+iN\big]=-a'^2 \left[(1+b^2)k^2-e^2\Phi^2\right]\left[(1-g)k^2-e^2\Phi^2\right],
\ee
where
\be
a'= \frac 14\left[4e^2\Phi^2-(4+ b^2)k^2\right]
\ee
and
\be
g= -\frac{e^{2}\Phi^{2}}{k^{2}-\frac{\lambda\Phi^{2}}{{6}}},
\ee
in which again we adopted the Feynman gauge and the limit $k_\mu k_\nu \to \eta_{\mu \nu} k^2/4$ was taken.

The effective potential is, thus, given by
\bq
\Gamma(\Phi)&=&-\frac{i}{2}\int \frac{d^{d}k}{(2\pi)^{d}}\left\{\ln\Big[\Big({k^{2}-\frac{\lambda}{6}\Phi^{2}}\Big)\Big({k^{2}-\frac{\lambda}{2}\Phi^{2}}\Big)\Big] \right. \nonumber \\
&+&\left.\ln\left\{-a'^2 \left[(1+b^2)k^2-e^2\Phi^2\right]\left[(1-g)k^2-e^2\Phi^2\right]\right\}\right\}.
\eq
The first term is the same as in the calculation for the Carroll-Field-Jackiw model and does not depend on the Lorentz-violating vector. For the second term, we can write, in Euclidean space
\begin{eqnarray}\label{53}
\Gamma_2(\Phi)=\frac{1}{2}\int \frac{d^{d}k_{E}}{(2\pi)^{d}}\ln\left\{\frac{\Big(\frac{4+b^2}{4}{k^{2}_{E}+e^2\Phi^2\Big)^2
\Big({k^{2}_{E}+c_1 \Phi^2}\Big)\Big(k^{2}_{E}+c_2 \Phi^2}\Big)\Big((1+b^2)k_E^2+e^2\Phi^2\Big)}{\Big(k^{2}_{E}+\frac{\lambda \Phi^2}{6}\Big)}\right\},
\end{eqnarray}
with $c_1$ and $c_2$ already defined in the first model. With the help of the Dimensional Reduction, we obtain
\bq
&&\Gamma_2(\Phi)=-\frac{\Phi^4}{32 \pi^2}\left\{-\frac{\lambda e^2}{3}\left[\frac{1}{\epsilon}+\frac 12 \ln\left(\frac{\Phi^2}{\mu^2}\right)\right]
+ \frac {c_1^2}{2} \ln c_1 + \frac{c_2^2}{2} \ln c_2 -\frac{\lambda^2}{72} \ln\left( \frac{\lambda}{6}\right) \right. \nonumber \\
&&\left.+e^4\left\{\left[ \frac{32}{(4+b^2)^2}+\frac{1}{(1+b^2)^2}\right]\left[\frac{1}{\epsilon}+\frac 12 \ln\left(\frac{\Phi^2}{\mu^2}\right)\right]
+\frac{16}{(4+b^2)^2} \ln\left(\frac{4e^2}{4+b^2}\right) \right. \right. \nonumber \\
&&\left. \left.+  \frac{1}{2(1+b^2)^2}\ln \left(\frac{e^2}{1+b^2}\right)\right\}\right\} + \mbox{constant terms}.
\eq

Again, it is important to note that the result of Coleman and Weinberg \cite{Col-Wei} is recovered in the limit $b_\mu \to 0$.

\subsection{The model with Lorentz symmetry breaking in a scalar sector}

\subsubsection{Formulation of the model}
We now set $\xi=\alpha=0$ and preserve only the Lorentz-breaking tensor $\kappa^{\mu \nu}$. The Lagrangian density is then given by
\begin{equation}\label{29}
{\cal L}_3=-\frac{1}{4}F_{\mu \nu}F^{\mu \nu}+\kappa_{\mu \nu}(D^\mu\phi)^*D^\nu\phi -\frac{\lambda}{4!}(\phi^* \phi)^2.
\end{equation}

The quadratic part of the action of the present model after the decomposition of the complex scalar field in its real components and the introduction of the background field is written as
\begin{eqnarray}\label{31}
\int d^4x{\cal L}(\phi_{a},\Phi_{a},A^{\mu})&=&\int d^4xd^4y\Big\{\frac{1}{2}A^{\mu}(x)(i\Delta^{-1})_{\mu\nu}A^{\nu}(y) +\nonumber \\&&
\frac{1}{2}\phi_{a}(x)(i{\cal D}^{-1})^{ab}\phi_{b}(y)+A^{\mu}(x){\cal M}_{\mu a}\phi^{a}(y)\Big\},
\end{eqnarray}
with
\begin{eqnarray}\label{32}
(\Delta^{-1})^{\mu\beta}&=&-k^{2}\eta^{\mu\beta}+k^{\mu}k^{\beta}(1-\chi) +e^{2}\kappa^{\mu\beta}\Phi^{2};\nonumber \\
({\cal D}^{-1})^{ab}&=& \kappa_{\mu\beta}\delta^{ab}k^{\mu}k^{\beta}-\frac{\lambda}{3}\Phi^{a}\Phi^{b}-\frac{\lambda}{6}\delta^{ab}\Phi^{2};\nonumber \\
{\cal M}_{\mu a}&=& -ie\kappa_{\mu\beta}\epsilon_{ab}\Phi^{b}k^{\beta}.
\end{eqnarray}

The background tensor $\kappa_{\mu \nu}$ should be symmetric and converge to $\eta_{\mu \nu}$ in the limit in which the Lorentz symmetry is restored. We will use a convenient and simple form for this tensor, given by $\kappa_{\mu \nu}=\eta_{\mu \nu} + b_\mu b_\nu$ with $b_\mu$ a dimensionless Lorentz-breaking vector.

\subsubsection{The effective potential}

For this third model, we follow the same steps for the calculation of the effective potential, given, as before, as
\begin{eqnarray}
\label{theint}
\Gamma(\Phi)&=&-\frac{i}{2}\int \frac{d^{d}k}{(2\pi)^{d}}\ln \det(i{\cal D}^{-1})\nonumber \\&& - \frac{i}{2}\int \frac{d^{d}k}{(2\pi)^{d}}\ln \det\big[i\Delta^{-1}+iN\big].
\end{eqnarray}
However, now, even in the purely scalar sector, we will have modifications. First, we have
\begin{equation}
\det \left({\cal D}^{-1}\right)
=\frac{\bar \kappa^2}{16}\Bigg({k^2-\frac{2\lambda}{3 \bar \kappa}\Phi^{2}}\Bigg)\Bigg({k^2-\frac{2\lambda}{\bar\kappa}\Phi^{2}}\Bigg),
\end{equation}
where again we have used the limit, under integration, $k_{\mu}k_{\beta} \rightarrow \frac{\eta_{\mu\beta}k^{2}}{4}$ and $\bar\kappa=\kappa_{\mu}^{\mu}$ represents the trace of the background tensor $\kappa_{\mu\beta}$. We also will have
\begin{equation}\label{34}
N_{\mu\beta}= i\frac{4e^2\Phi^2}{\bar \kappa} \frac{\tilde k_\mu \tilde k_\beta}{k^2-\frac{2\lambda \Phi^2}{3 \bar \kappa}},
\end{equation}
with $\tilde k_\mu=\kappa_{\mu \nu}k^\nu$. So, we get
\begin{eqnarray}\label{35}
i(\Delta^{-1})_{\mu\beta}+i N_{\mu\beta}&=&-k^{2}\eta^{\mu\beta}+k^{\mu}k^{\beta}(1-\chi) +e^{2}\kappa^{\mu\beta}\Phi^{2}\nonumber \\
&& - \frac{4e^2\Phi^2}{\bar \kappa} \frac{\tilde k_\mu \tilde k_\beta}{k^2-\frac{2\lambda \Phi^2}{3 \bar \kappa}},
\end{eqnarray}
with the following result for the determinant in the Feynman gauge, after some manipulations and the use of the limit $k_\mu k_\nu \to k^2 \eta_{\mu \nu}/4$,
\bq
\label{det}
\det\big[i\Delta^{-1}+iN\big]&=&-\frac 14 \left[-k^2+e^2\Phi^2\right]^2 \left\{\left[-k^2+e^2\Phi^2\right]
\left[\left(g(b^2+2)^2-4\right)k^2+4(1+b^2)e^2\Phi^2\right] \right.\nonumber \\
&+& \left. 3gb^2e^2\Phi^2 k^2\right\},
\eq
with
\be
g=- \frac{4e^2\Phi^2}{\bar \kappa} \frac{1}{k^2-\frac{2\lambda \Phi^2}{3 \bar \kappa}}.
\ee
The integral (\ref{theint}) with the determinant given by (\ref{det}) apparently furnishes a complicated result in its explicit form. However, its general form can be obtained. If one considers $\lambda=\alpha e^2$, with $\alpha$ some number, it is clear that, by dimensional reasons, the renormalized result for the effective potential will look like
\be
\label{res}
\Gamma_2(\Phi)=e^4\Phi^4(a_1+a_2\ln\frac{\Phi^2}{\mu^2}),
\ee
where $a_1$ and $a_2$ are some finite constants depending on $\bar{\kappa}$. Also,  we note that, for light-like $b_{\mu}$, one has
\bq
\det\big[i\Delta^{-1}+iN\big]&=&- \left[-k^2+e^2\Phi^2\right]^3
\left[(g-1)k^2+e^2\Phi^2\right],
\eq
 with $\bar{\kappa}$ in this case is simply 4, and in this case we evidently reproduce the Lorentz invariant result.

It is interesting to comment on the general dependence of the effective potential on the Lorentz-breaking parameters. It was shown in \cite{Carvalho} (and in supersymmetric case, in \cite{ours}) that in theories where the modified Lorentz-breaking kinetic term of the scalar field is proportional to $(\eta_{\mu\nu}+\tilde{\kappa}_{\mu\nu})\pa^{\mu}\phi\pa^{\nu}\phi$, with $\kappa_{\mu\nu}=\eta_{\mu\nu}+\tilde{\kappa}_{\mu\nu}$, the $L$-loop correction is proportional to $\det^{-L/2}(1+\tilde{\kappa})$, hence, the constants $a_1$ and $a_2$ from (\ref{res}) will be proportional to $\det^{-1/2}(1+\tilde{\kappa})$. 

This conclusion allows us to make a final estimation of Lorentz-breaking impacts for the effective potential. Indeed, we have argued that the effective potential will be corrected by the multiplier $\det^{-1/2}(1+\tilde{\kappa})$, which, for $|\tilde{\kappa}_{\mu\nu}|\ll 1$, can be represented as 
$1-\frac{1}{2}\tilde{\kappa}_{\mu}^{\mu}$. Therefore, we can see that the Lorentz-breaking modifications to the effective potential will be of the order of $\tilde{\kappa}_{\mu}^{\mu}$, which, following \cite{datatables}, has the order $10^{-27}$. Hence, already the LV leading order contribution will be very tiny, differing from the usual result by 27 orders.

\section{Conclusions}

In this paper, we studied how the one-loop effective potential is modified by Lorentz-symmetry violation in some extended models. We note that it is the first calculation of this type since, up to now, the effective potential in a Lorentz-breaking case has been evaluated only in a purely scalar theory \cite{aether}. First, we considered a massless Yukawa-like model and, then, some examples of Lorentz-breaking scalar QED. The one-loop corrections to the effective potential in all the cases we studied depend on the background tensors responsible for the Lorentz-symmetry violation, but converge to the known results in the limit Lorentz-symmetry is restored. More interesting is the fact that this limit is recovered even in the presence of the Lorentz-breaking vectors if they are light-like. Particularly, for the massless Yukawa-like model, the effective potential vanishes if the four-vector $b_\mu$ is light-like.

It is well-known from the paper \cite{Jackiw} that the effective potential in the traditional massless scalar QED is gauge dependent. This is explained by the fact that the shift in the scalar field performed in the process of calculation of effective potential induces a mass for the gauge field. Despite this, we adopted in our calculation the Feynman gauge, since we are interested here in observing the dependence of the results on the Lorentz-breaking parameters. One of the physical aspects to be discussed further is the possible dependence of the induced masses on the background tensors responsible for the Lorentz violation. It would be interesting for a future investigation to check if there exists a particular gauge in which this dependence of physical results on Lorentz breaking parameters is removed.

{\bf Acknowledgements.} This work was partially supported by Conselho
Nacional de Desenvolvimento Cient\'{\i}fico e Tecnol\'{o}gico (CNPq). The work by A. Yu. P. has been supported by the
CNPq project No. 303783/2015-0.







\begin{thebibliography}{99}


\bibitem{Kostel} V. A. Kostelecky, Phys. Rev. D69, 105009 (2004), hep-th/0312310.

\bibitem{class} H. Belich, T. Costa-Soares, M. M. Ferreira Jr., J. A. Helayel-Neto, Eur. Phys. J. C41, 421 (2005), hep-th/0410104; Eur. Phys. J. C42, 127 (2005), hep-th/0411151; H. Belich, T. Costa-Soares, M. M. Ferreira Jr., J. A. Helayel-Neto, F. M. O. Moucherek, Phys. Rev. D74, 065009 (2006), hep-th/0604149; H. Belich, L. P. Colatto, T. Costa-Soares, J. A. Helayel-Neto, M. T. D. Orlando, Eur. Phys. J. C62, 425 (2009), arXiv: 0806.1253; R. Casana, M. M. Ferreira, C. Santos, Phys. Rev. D78, 105014 (2008), arXiv: 0810.2817; R. Casana, M. M. Ferreira, J. S. Rodrigues, M. R. O. Silva, Phys. Rev. D80, 085026 (2009), arXiv: 0907.1924;  R. Casana, A. R. Gomes, M. M. Ferreira, P. R. D. Pinheiro,
Phys. Rev. D80, 125040 (2009), arXiv: 0909.0544;
F. R. Klinkhammer, M. Schreck, Nucl. Phys. B848, 90 (2011), arXiv: 1011.4258; Nucl. Phys. B856, 666 (2012), arXiv: 1110.4101; M. Schreck, Phys. Rev. D86, 065038 (2012), arXiv: 1111.4182;
R. Casana, E. S. Carvalho, M. M. Ferreira,
Phys. Rev. D84, 045008 (2011), arXiv: 1107.2664; R. Casana, M. M. Ferreira, R. P. M. Moreira,
Phys. Rev. D84, 125014 (2011), arXiv: 1108.6193;	
A. P. Ba\^eta Scarpelli, M. Gomes, A. Yu. Petrov, A.J. da Silva, Phys.Rev. D93, 2, 025010 (2016), arXiv:1509.05352. 

\bibitem{CMR} C. M. Reyes, L. F. Urrutia, J. D. Vergara, Phys.Lett. B675, 336 (2009), arXiv: 0810.4346; Phys. Rev. D78, 125011 (2008), arXiv: 0810.5379; C. Marat Reyes,  Phys. Rev. D80, 105008 (2009), arXiv: 0901.1341; Phys. Rev. D82, 125036 (2010), arXiv: 1011.2971; Phys. Rev. D87, 125028 (2013), arXiv: 1307.5340; M. Maniatis, C. Marat Reyes, Phys. Rev. D89, 056009 (2014), arXiv: 1401.3752.

\bibitem{JK} R. Jackiw, V. A. Kostelecky, Phys. Rev. Lett. 82, 3572 (1999), hep-ph/9901358.

\bibitem{CFJ} S. Carroll, G. B. Field, R. Jackiw, Phys. Rev. Lett. 41, 1231 (1990).



\bibitem{ourHD} T. Mariz, Phys. Rev. D83, 045018 (2011), arXiv: 1010.5013; T. Mariz, J. R. Nascimento, A. Yu. Petrov, Phys. Rev. D85, 125003 (2012), arXiv: 1111.0198; J. Leite, T. Mariz, W. Serafim, J. Phys. G40, 075003 (2013).

\bibitem{Col-Wei} S. Coleman, E. Weinberg, Phys. Rev. D7, 1888 (1973).

\bibitem{aether} M. Gomes, J. R. Nascimento, A. Yu. Petrov, A. J. da Silva, Phys. Rev. D81, 045018 (2010), arXiv: 0911.3548.


%
\bibitem{Collins} J.~Collins, A.~Perez, D.~Sudarsky, L.~Urrutia and H.~Vucetich,
  Phys.\ Rev.\ Lett.\  {\bf 93}, 191301 (2004),
  gr-qc/0403053.

\bibitem{Reyes}  C.~M.~Reyes, S.~Ossandon and C.~Reyes,
  Phys.\ Lett.\ B {\bf 746}, 190 (2015),
  arXiv: 1409.0508.
	
\bibitem{Carvalho} P. R. S. Carvalho, M. I. Sena-Junior, Ann. Phys. 387, 290 (2017); Eur. Phys. J. C77, 753 (2017).	

\bibitem{ours} C. F. Farias, A. C. Lehum, J. R. Nascimento, A. Yu. Petrov, Phys. Rev. D86, 065035 (2012), arXiv: 1206.4508; A. C. Lehum, J. R. Nascimento, A. Yu. Petrov, A. J. da Silva, Phys. Rev. D88, 045022 (2013), arXiv: 1305.1812.

\bibitem{Jackiw} R. Jackiw, Phys. Rev. D9, 1686 (1974).

\bibitem{datatables} V. A. Kostelecky, N. Russell, Rev. Mod. Phys. 83, 11 (2011), arXiv: 0801.0287.

\end{thebibliography}
\end{document}